\documentclass[iop]{emulateapj}
\usepackage{graphicx}  
\usepackage{natbib}

\begin{document}

\title[A search for consistent jet and disk rotation signatures in RY Tau]{A search for consistent jet and disk rotation signatures in RY Tau}

\author{Deirdre Coffey}
\address{School of Physics, University College Dublin, Belfield, Dublin 4, Ireland.}

\author{Catherine Dougados}
\address{UMI LFCA Universidad de Chile}
\address{I.P.A.G. (UMR 5274), BP 53, F-38041 Grenoble C\'{e}dex 9, France.}

\author{Sylvie Cabrit} 
\address{LERMA, Observatoire de Paris, UMR 8112 du CNRS, ENS, UPMC, UCP, 61 Av. de l'Observatoire, F-75014 Paris, France.}
\address{I.P.A.G. (UMR 5274), BP 53, F-38041 Grenoble C\'{e}dex 9, France.}

\author{Jerome Pety} 
\address{I.R.A.M., 300 rue de la Piscine, Domaine Universitaire, 38406 Saint Martin d'Hres, France.} 
\address{LERMA, Observatoire de Paris, UMR 8112 du CNRS, ENS, UPMC, UCP, 61 Av. de l'Observatoire, F-75014 Paris.}

\author{Francesca Bacciotti}
\address{I.N.A.F., Osservatorio Astrofisico di Arcetri, Largo E. Fermi 5,  50125 Florence, Italy.}


\begin{abstract} 

We present a radial velocity study of the RY Tau jet-disk system, designed to determine whether a transfer of angular momentum from disk to jet can be observed. Many recent studies report on the rotation of T Tauri disks, and on what may be a signature of T Tauri jet rotation. However, due to observational difficulties, few studies report on both disk and jet within the same system to establish if the senses of rotation match and hence can be interpreted as a transfer of angular momentum. We report a clear signature of Keplerian rotation in the RY Tau disk, based on Plateau de Bure observations. We also report on the transverse radial velocity profile of the RY Tau jet close to the star. We identify two distinct profile shapes: a v-shape which appears near jet shock positions, and a flat profile which appears between shocks. We do not detect a rotation signature above 3 sigma uncertainty in any of our transverse cuts of the jet. Nevertheless, if the jet is currently in steady-state, the errors themselves provide a valuable upper limit on the jet toroidal velocity of 10 km~s$^{-1}$, implying a launch radius of $\le$ 0.45 AU. However, possible contamination of jet kinematics, via shocks or precession, prevents any firm constraint on the jet launch point, since most of its angular momentum could be stored in magnetic form rather than in rotation of matter. 

\end{abstract}

\maketitle 

\vspace{2pc} 

\section{Introduction} 
\label{intro} 

The origin of protostellar jets, their role in angular momentum extraction and their impact on protoplanetary disks are still critical open issues in star formation. The MHD ejection mechanism originally proposed by \citep{Blandford1982} is currently the favoured process for jet generation as it can naturally account for both high jet speeds and high degrees of collimation close to the source. MHD jets from young accreting stars may have three possible components (see review by \citet{Ferreira2006} and references therein):  a strong pressure-driven stellar wind (SW); sporadic wide-angle magnetospheric ejections (MEs) caused by the interaction of the stellar magnetosphere with the inner disk edge and a self-collimated centrifugally-driven disk wind either originating from the co-rotation radius (X-wind solution) or from a wider range of disk radii (extended disk winds).  MEs and SWs currently appear as the best candidates to spin-down young accreting stars below break-up. If extended disk-winds are present in young stars, this would imply that the inner astronomical units of proto-planetary disks are more strongly magnetised than assumed so far, which would have strong implications for angular momentum transfer and planetary formation and migration models (see reviews by \citealp{Turner2014} and \citealp{Baruteau2014}). 

Over the last decade, both semi-analytical modelling and numerical simulations have been extensively used to investigate how the mass-loading, kinematics and collimation properties of disk winds are contribed but the detailed physics of the underlying accretion disk (magnetic field intensity and topology, thermodynamics, resistivity etc (\citealp{Casse2000, Fendt2002}, 2006, \citealp{Zanni2007, Murphy2010, Sheikhnezami2012, Stepanovs2014}). A direct comparison of jet observations with model outputs remains difficult because current numerical works remain biased towards small magnetic lever arm solutions, probably due to numerical diffusivity \cite{Murphy2010}. In addition, a proper treatment of non-equilibrium temperature and ionisation is required, which can completely change the emission line profiles for the underlying dynamics. So far, this has only been done by \citet{Te&scedil;ileanu2014}, and only for simulations treating the disk as a platform, which are not dynamically self-consistent. Conversely, semi-analytical solutions remain limited by the self-similar assumption and the lack of proper inner and outer boundary conditions. 

An alternative approach, which we adopt here, is that of relying on general conservation principles of any steady, axisymmetric, cold magneto-centrifugal disk wind solution after it is launched and in ideal MHD, namely the conservation of total (matter plus magnetic) specific energy and angular momentum along each magnetic surface. \citet{Anderson2003} showed that a combination of these two invariants allows us to get rid of the non-observable magnetic terms and constrain the launching radius of the disk wind streamline independently of any specific MHD model, simply from the observed jet rotational and poloidal velocity. \citet{Ferreira2006} showed that a lower limit on the disk wind magnetic lever arm can also be obtained in this way. 

In recent years, pushing the limits of observational resolution has revealed a number of exciting detections of gradients in the radial velocity profile {\em across} jets from T~Tauri stars (\citealp{Bacciotti2002}, \citealp{Coffey2004}, 2007, \citealp{Woitas2005}). These may be interpreted as a signature of jet rotation about its symmetry axis, thereby representing the long-awaited observational support of the theory that MHD jets extract angular momentum from star-disk systems. Under the assumption of steady mass-loss, deriving jet angular momentum has the potential to constrain the launching radius (\citealp{Bacciotti2002}; \citealp{Anderson2003}). If the observed transverse velocity gradients are due to rotation, then they imply launching radii of 0.1-3 AU \citep{Ferreira2006}. However, the possibility that we are indeed observing jet rotation in pre-mainsequence systems is undergoing active debate. Alternative interpretations include asymmetric shocking and/or jet precession (e.g. \citealp{Soker2005}, \citealp{Cerqueira2006}, \citealp{Correia2009}). It is therefore important to perform new checks on the rotation interpretation. 

First, to ensure that we probe the kinematics of the launching process, the jet must be observed as close as possible to the star where any evidence of angular momentum transfer is still preserved, rather than far from the star by which time the internal shocks, or interaction with the ambient medium, may have disrupted any rotation signature intrinsic to the launch mechanism. Second, we must check that rotation signatures are consistent at different positions along the jet (in various tracers and epochs). This has been checked in only three studies so far:  an optical study of DG Tau \citep{Bacciotti2002} and RW Aur \citep{Woitas2005} where consistency was found; a UV study of DG Tau and RW Aur where consistency was found in the former but not the latter (\citealp{Coffey2007}, \citealp{Coffey2012}); and a near infrared study of DG Tau \citep{White2014} where systematic transverse velocity gradients could not be identified). Third, we must check that the jet Doppler gradient matches the direction of the disk Doppler gradient. Detecting both jet and disk gradients in the same protostellar system remains challenging, with the few successes to-date leading to no clear conclusion. Some studies have reported a match in gradient direction, while others claim no match and others still claim no detection of gradients in the jet. Studies close to the jet footpoint usually require the jet to be optically visible, and so Class II T Tauri stars become the targets. In these cases, matching jet gradients within 100 AU are reported for DG~Tau \citep{Testi2002, Bacciotti2002} and CW Tau (Dougados et al. in prep.), but no clear detection of jet rotation is reported for HH~30 (\citealp{Coffey2007}, \citealp{Pety2006}, \citealp{Guilloteau2008}), and mis-matched gradients are reported for RW~Aur (\citealp{Cabrit2006, Coffey2012}). 

Persevering with this investigation, our latest study examines the jet from RY Tau, a 2 M$_\sun$ T Tauri star of spectral type F8 located at 140 pc. An blueshifted atomic jet at position angle 294$\degr$ has been detected both on large scales in H$\alpha$ \citep{St-Onge2008} and on small scales in [\ion{O}{1} \citep{Agra-Amboage2009}. The disk has been resolved in millimetric continuum emission \citep{Isella2010}. These observations constrain the inclination of the disk rotation axis to the line of sight between 65-75 $\degr$, consistent with the UX Ori behavior.  We present a high resolution spectro-imaging study with GEMINI/NIFS+Altair of the RY Tau jet in the [Fe {\sc ii}] 1.64 $\mu$m emission line. We also present a study of the disk kinematics conducted with IRAM/PbBI in millimetric CO lines. Together these data sets allow us to search for velocity gradients both in the jet and disk, and simultaneously perform the two consistency checks for jet rotation signatures discussed before. Furthermore, the imaging capabilities of NIFS/GEMINI provides an insight into the impact of variability within the jet on the derivation of rotation signatures. Our observations are described in section \ref{obs}, while the method of analysis and results are detailed in sections \ref{calib}, \ref{analysis} and \ref{results}. In section \ref{discussion}, we discuss the measurements in the context of the magneto-centrifugal acceleration process, and we summarise our conclusions in section \ref{conclusions}.

\section{Observations}
\label{obs} 

To observe the outflow from RY Tau (R=9.67), GEMINI/NIFS+Altair was used to obtain AO-corrected (with natural guide star) IFU data of the approaching jet (PA=294$\degr$), with the H-band filter/grating at slit/slice position angles both parallel (24$^\circ$) and anti-parallel (204$^\circ$) to the jet axis (Program ID: GN-2009B-Q-43). This configuration positioned the IFU 0.1$\arcsec$$\times$0.043$\arcsec$ spaxels with the smallest sampling in the direction perpendicular to the jet axis. Additional arc lamp exposures were taken during observations to ensure high accuracy wavelength calibration. Velocity resolution of the instrument was 57 km~s$^{-1}$, but for good signal-to-noise we could achieve higher velocity precision in emission line velocity centroids, via Gaussian fitting (section \ref{analysis}). Spatial resolution with AO correction (measured as the FWHM of the stellar PSF core) in the direction across the jet is 0$\farcs$12 (17 AU), and along the jet is 0$\farcs$2 (28 AU). To retrieve the faint jet emission {\it close} to the star (where we wish to probe for jet rotation), accurate subtraction of the stellar continuum was required. This necessitated inclusion of the star within the 3$\arcsec$$\times$3$\arcsec$ field of view, and hence a reduction in the exposure time (to 10 s) with a consequent increase in the number of co-added frames, in order to achieve a jet signal-to-noise of typically 35. The star was offset from the centre of the field so as to observe as much of the jet as possible while still ensuring coverage of the entire stellar PSF. With this setup, the IFU captured the jet out to 1.7\arcsec~from the star. One third of the time was spent on sky exposures, via the standard ABA nodding technique. Unfortunately, some of the observations with the anti-parallel slit show that the field of view (FOV) was re-centered on the star, rather than offsetting to the jet, causing a loss of information on the outer jet emission. Therefore, these observations were not included in the co-adds and hence the jet was not observed with the same signal-to-noise at each position angle. It subsequently transpired that the anti-parallel slit data could not be used as it did not reach sufficiently high signal-to-noise. 

To observe the RY Tau disk, $^{12}$CO J=2-1 observations of the system were performed with the IRAM Plateau de Bure Interferometer on 2007 December 18 using the extended C configuration with 6 antennas and baselines ranging from 24m to 176m (project Q010). The 1 mm receivers were tuned single sideband at 230.538 GHz. One correlator band of 40 MHz was centered on the  $^{12}$CO J = 2-1, implying a channel spacing of 78 kHz or 0.1 km~s$^{-1}$. The total telescope time, including phase and amplitude calibrators,  amounted to  24h with 6 antennas.  

\begin{table*}
\caption{\label{tab1}
GEMINI/NIFS+Altair observations of the RY Tau blue-shifted jet. 
} 
\begin{center}
\begin{tabular}{@{}*{4}{l}}
\tableline
Observation~Date & Slitlets PA & Co-added Exposure Times  & Total Time \cr 
 & ($^\circ$) & (s)  & (hrs) \cr 
\tableline
14 Dec 2009 & 24  & 3060, 1020, 1530, 2040 & 2.125 \cr
15 Dec 2009 & 204  & 4080, 3060 &1.983 \cr
24 Dec 2009 & 24  & 2040 &  0.567  \cr
\tableline
\end{tabular}
\end{center}
\end{table*}

\section{Data Reduction}
\label{calib}

The NIFS data were processed through the standard GEMINI calibration pipeline, but using the additional arc lamp files to improve wavelength calibration. It was not necessary to perform telluric subtraction during the pipeline, because subsequent continuum subtraction took into account the telluric features. The output IFU cubes for a given position angle were then combined, and continuum subtraction performed as follows. We selected a high signal-to-noise reference photospheric spectrum free of jet emission. This reference spectrum was then scaled down to the level of continuum in a region near the emission line of interest, and was subtracted from the jet spectrum. The residual emission line profile is not only free of photospheric absorption lines but also of telluric features. Note, this procedure only fully corrects the telluric absorption formed against the stellar continuum. It does not correct for telluric absorption against the extended [\ion{Fe}{2}] emission. However, we checked that this is a small effect (less than 10\% of the total [\ion{Fe}{2}] emission intensity) (see \citet{Agra-Amboage2011} for further details). In our data, the brightest jet emission line was [\ion{Fe}{2}] at 1.644 \micron. Continuum subtraction was conducted with due care in this region of the spectrum, and was sufficiently accurate to allow retrieval of jet spectra as close to the jet base as 0$\farcs$2. Observations of the sky were processed through the same pipeline as the science observations. In this way, we ensure that the sky observations carry the same velocity accuracy as the science data. We can then profile fit the sky lines to determine the velocity calibration error, which will be the same as the science data. 

The Plateau de Bure data were reduced using the GILDAS\footnotemark[1] software. Standard calibration methods, using close calibrators, were applied. The phase and amplitude calibrators 0528+134, J0418+380 were observed along with 3C454.3 as radio-frequency bandpass calibrator. The low quality data were filtered out. In particular, data with phase noise rms above 50$\degr$ were not used in subsequent data analysis. $^{12}$CO J=2-1 images were produced using the CLEAN algorithm. The clean beam size is 1$\farcs$4500 $\times$ 1$\farcs$0100 at PA=41$\degr$. Velocities are expressed in the LSR frame with a systemic velocity of 7.8 km~s$^{-1}$. 

\footnotetext[1]{See http://www.iram.fr/IRAMFR/GILDAS for more information on GILDAS software.}

\section{Data Analysis}
\label{analysis} 

The individual NIFS spectra showed that the [\ion{Fe}{2}] line was well fitted with a single Gaussian. Emission line profile fitting was conducted at every position along and across the jet. The resulting velocities were corrected by -17 km~s$^{-1}$ to the LSR velocity frame. 

The overall error in centroid velocity was determined by measuring the velocities of the OH skylines. To do this, four of the brightest sky lines were selected (at 1.60307, 1.61283, 1.66924 and 1.69035 \micron) based on their close proximity to the [\ion{Fe}{2}] 1.644 \micron ~jet emission line. The velocities were measured in each spaxel, and their RMS was calculated separately for each slicing mirror, as this is the direction across the jet. Hence, we could identify wavelength calibration drifts which may affect our measurement of the transverse radial velocity profile of the jet. 

Each RMS value comprises a centroid fitting error and a velocity calibration error, only the former of which depends on signal-to-noise. That is, the theoretical standard deviation, $\sigma$, in estimating the centroid of a Gaussian distribution is obtained by  \citep{Porter2004}: 
\begin{equation}\label{eq1} 
\sigma = \frac{FWHM}{2.35~SNR} 
\end{equation} 
where $FWHM$ in this case is measured in the spectral direction, and $SNR$ is the signal-to-noise ratio at the Gaussian peak. 

\begin{figure} 
\begin{center} 
\includegraphics[ width=1\columnwidth]{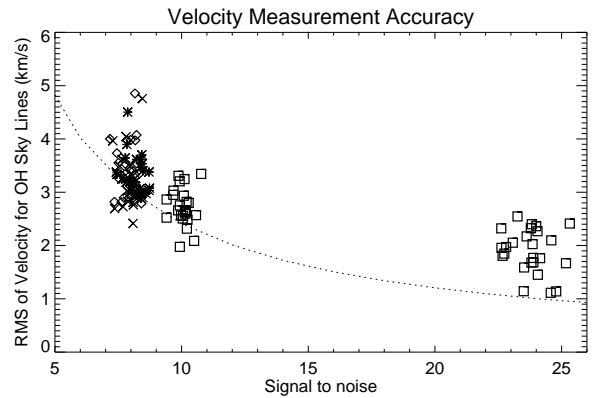} 
\caption{RMS velocities plotted for each slicing mirror (slitlet) of each skyline. Each skyline is represented by a different symbol. The brightest skyline (snr=10) was then binned ($\times$4) along the slitlet and the improved RMS was plotted (see points at snr=25). 
The dashed line represents the theoretical 1$\sigma$ error on the Gaussian centroid fit, calculated according to equation \ref{eq1}.  
\label{fig1} 
} 
\end{center} 
\end{figure}

For each skyline, the RMS error was plotted against the median signal-to-noise, figure \ref{fig1}, and a curve marking the theoretical fitting error (according to equation \ref{eq1}) was added for comparison. Binning ($\times$4 along the slitlet) of the brightest skyline (snr=10) allowed extrapolation of the error to a signal-to-noise comparable to that of the jet emission (i.e. SNR=25). Subtracting in quadrature the theoretical fitting error (curve) from the total error (data points) allows extraction of the velocity calibration error, $\sigma_{cal}$, which in our case averages at 1.6 km s$^{-1}$ (RMS). Using this value, we can now calculate the total error in centroid velocity for jet emission of any signal-to-noise, by simply adding the velocity calibration error (i.e. $\sigma_{cal}$) in quadrature to the fitting error (i.e. the output of equation \ref{eq1}). 

For example, for a jet signal-to-noise of 25, and given the spectral resolution of NIFS (FWHM=57 km s$^{-1}$), the 1 $\sigma$ error on the velocity centroid fitting is 1 km s$^{-1}$ (in line with the lower limit on the sky line RMS, figure \ref{fig1}). Adding this in quadrature to the velocity calibration error of 1.6 km s$^{-1}$, yields an overall 3 sigma velocity error of 5.7 km s$^{-1}$. Multiplying by $\sqrt2$, we obtain a 3 $\sigma$ error on differences in Doppler shift of 8 km s$^{-1}$. Note that this is a {\it lower} limit, based on a maximum jet signal-to-noise of 25-30. For the jet borders, where signal-to-noise can drop to 10-15, this error grows to 10-13 km s$^{-1}$. Previously published jet rotation indications are typically on the order of 1-15 km s$^{-1}$.  So, in spite of our strategy to increase velocity accuracy by including  wavelength calibration arc lamp exposures during the observations, we still find ourselves operating at the very limit of the instrument's ability to detect differences in Doppler shift across the jet. 

Finally, note that the calibration error, $\sigma_{cal}$, includes random spaxel to spaxel fluctuations and systematic drifts.  We examined such drifts and no systematic trends in skyline velocity measurements above 3$\sigma_{cal} > $5 km s$^{-1}$ were identified across the slitlets/slicing mirrors (i.e. along the jet). Along the slitlets (i.e. across the jet), a slight trend was identified. However, the gradient did not always have the same slope, and the amplitude was low ($\sim$ 2-3 km s$^{-1}$), hence it remains well within the overall three sigma error bars. Furthermore, it is possible that a rotation signature could be mimicked by a contribution from an uneven slit illumination, due to the strong gradient in brightness within a given slitlet caused by the continuum of the star. (I.e. when we compute the transverse radial velocity differences, we take into account the fact that the jet axis is wiggling such that uneven slit illumination effects will introduce a spurious value, even if the PSF is exactly symmetric with respect to the axis of the detector which it is not). We estimated the amplitude of uneven slit illumination effects on transverse radial velocity measurements using the procedure outlined in \citet{Agra-Amboage2014}. This was found to introduce an effect on the order of 0.5 km s$^{-1}$ which was deemed negligible in view of the overall error bars. 

\section{Results} 
\label{results} 

While the NIFS and PDBI datasets presents a wealth of information on the jet and disk morphology and kinematics, we choose to restrict our current focus to the case for jet rotation. Two separate more comprehensive studies of jet and disk are to follow (Coffey et al. in preparation; Dougados et al. in preparation), with just a brief commentary on the jet given below by way of introduction. 

\subsection{Jet Morphology and Kinematics} 
\label{MnK}

\begin{figure}
\begin{center}
\includegraphics[ angle=-90,width=1.0\columnwidth]{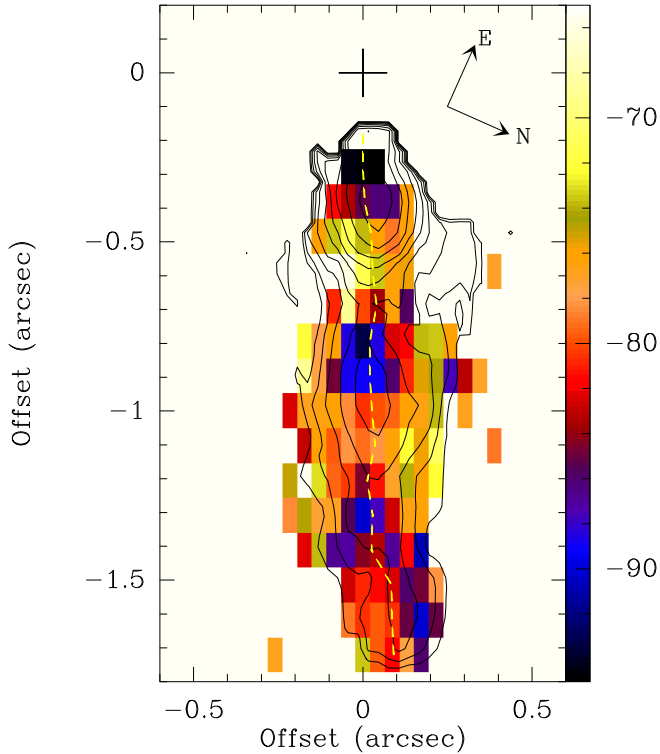} 
\caption{Radial velocity map of the RY Tau jet in [\ion{Fe}{2}]  1.644 $\micron$ emission. The map is overlaid with intensity contours, and the position of the jet axis (dashed line). The approaching jet PA = 294 $\degr$. 
\label{fig2}
}
\end{center}
\end{figure}

It is interesting to note the general morphology and kinematics of the jet, as revealed by the [\ion{Fe}{2}] 1.644$\micron$ emission line. A contour plot of the jet intensity, figure~\ref{fig2}, shows peaks are located at positions around -0$\farcs$2 to -0$\farcs$5 and -0$\farcs$8 to -1$\farcs$1, which indicates a clear presence of internal jet shocks (or knots) along the flow direction. The jet axis is over-plotted as a dashed central curve. It was determined by gaussian fitting the intensity profile across the jet image. At each position in the image, the radial velocity was obtained by Gaussian fitting the spectral emission peak where signal-to-noise $\ge$ 5. Figure~\ref{fig2} shows how the contours of the jet intensity relate to the velocity colour map. Note how peaks in intensity contours coincide with local maxima in the on-axis velocity. In the second knot (at -0$\farcs$8), it is clear that the velocity peak slightly precedes the intensity peak, which supports the presumption that we are observing emission from shock-excited gas. 

\begin{figure*}
\begin{center}
\includegraphics[ width=1.5\columnwidth]{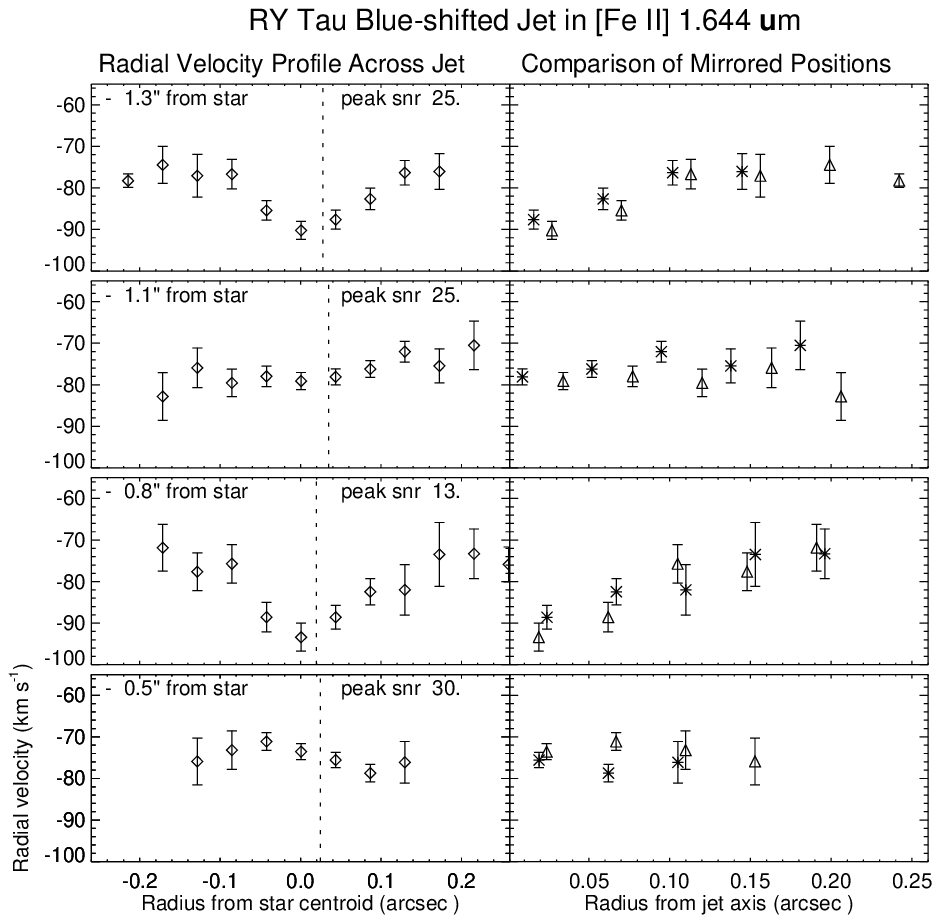}
\caption{Left panels: Transverse velocity profiles along the jet at various positions: -0.5$\arcsec$, -0.8$\arcsec$, -1.1$\arcsec$  and -1.3$\arcsec$. The position of the jet axis is indicated by the dashed line. Right panels: comparison of centroid velocities on either side of the jet axis. Error bars are at the 1$\sigma$ level. 
\label{fig3}
}
\end{center}
\end{figure*}

\begin{figure*}
\begin{center}
\includegraphics{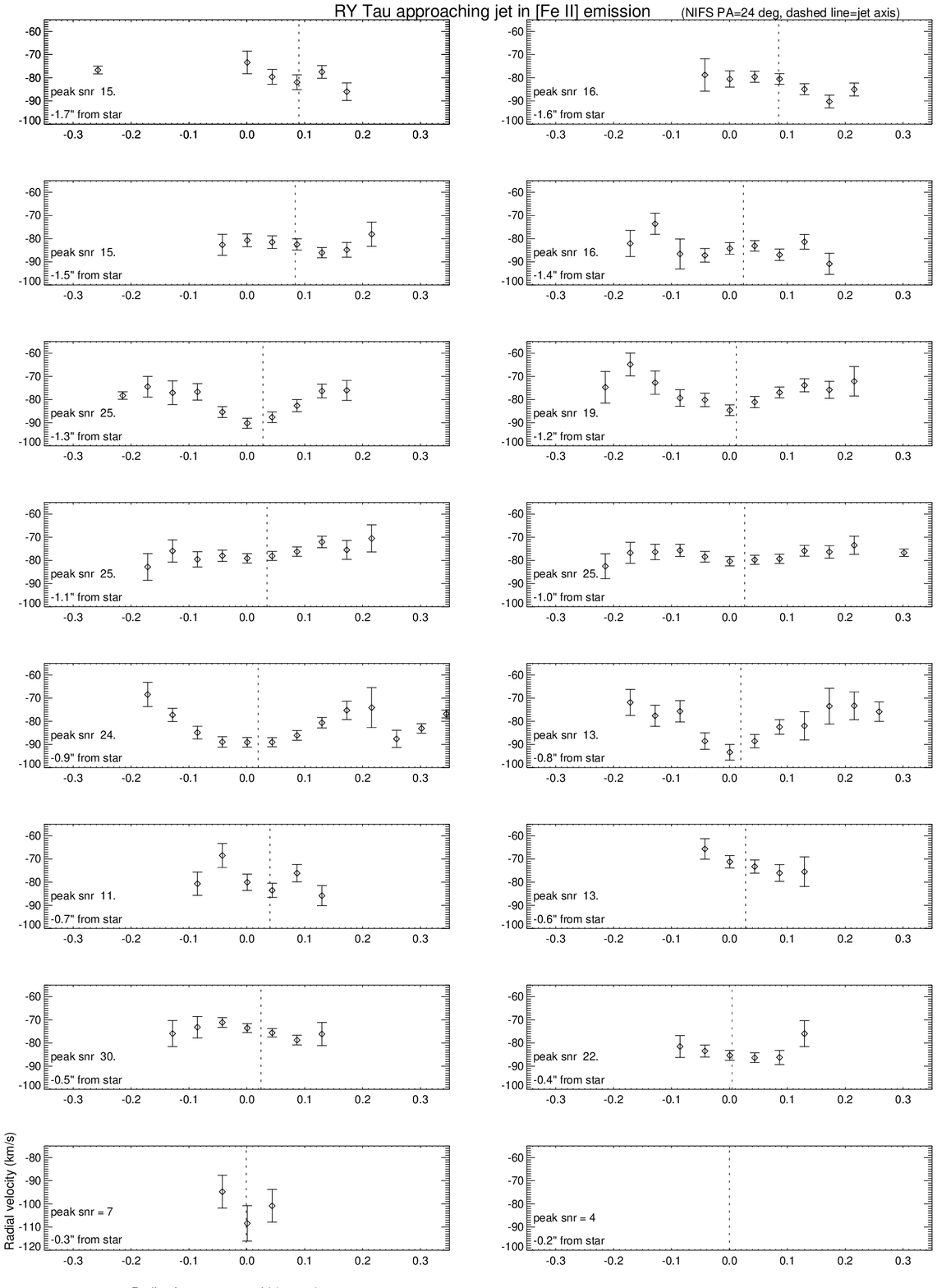} 
\caption{Radial velocity profiles across the RY Tau jet in [\ion{Fe}{2}]  1.644 $\micron$ emission, for NIFS PA=24 degrees. Velocities were determined via Gaussian fits to emission of signal-to-noise great than 5. Error bars are at the 1 $\sigma$ level. The dashed line represents the jet axis, as determined by Gaussian fits to the image intensity. Close to the star, the noise from PSF subtraction increases which reduces the jet SNR. 
\label{fig4}
}
\end{center} 
\end{figure*}

Recall that the spatial resolution (measured via the stellar PSF) in the direction across the jet is 0$\farcs$12 (17 AU), and along the jet is 0$\farcs$2 (28 AU). Meanwhile, we measure a jet FWHM of 0$\farcs$18 (25 AU), at -0$\farcs$5 (70 AU) projected distance from the star. Therefore, the jet width is resolved in this region close to the star. Figure~\ref{fig3} shows the transverse radial velocity profile, at four sample positions along the jet. (Figure \ref{fig4} shows the radial velocity profile across the jet at {\it all} positions along the jet.) Along the jet, we can distinguish two shapes of transverse radial velocity profile. A v-shaped profile appears where the jet velocity peaks on-axis, at positions -0$\farcs$3, -0$\farcs$8 to -0$\farcs$9 and -1$\farcs$2 to -1$\farcs$3. Again, this seems to coincide with intensity peaks of the shocked jet material. Between shocks, we see a flat profile, at positions -0$\farcs$4 to -0$\farcs$6, -1$\farcs$0 to -1$\farcs$1 and -1$\farcs$4 to -1$\farcs$5. 

To add to the complexity, we also see signs of wiggling of the jet as it propagates, as indicated by the dashed curve in figure~\ref{fig2} and the dashed central line of figures ~\ref{fig3} and \ref{fig4}. Indeed, the jet wiggle, which becomes apparent after -0$\farcs$5, implies that any kinematic record of the jet launch mechanism is probably contaminated beyond this point. Nevertheless, in this case, its amplitude is small and we take it into account when considering transverse kinematic signatures which may represent jet rotation. Although the wiggle seems to cause the jet to veer to the side at -1$\farcs$25, the jet is in fact opening out into a v-shape, the left arm of which falls below the 5 sigma level. 

Our data show that radial velocity profiles can be dominated by shock kinematics and jet wiggle, rather than any record of the jet launch mechanism, once the jet has propagated beyond about -0$\farcs$5. This highlights the need to examine kinematics as close to the source as possible, in order to reduce the impact of these effects. 

\subsection{Jet Rotation} 

If the jet is rotating, we should observe a difference in the Doppler shift between one side of the jet axis and the other. Examining radial velocity measurements across the jet, we do not see any clear sign of jet rotation at the 3$\sigma$ level. In the right-hand panels of figure \ref{fig3}, we compare the radial velocities at symmetric positions about the jet axis (taking account of the jet wiggle). We do not find any clear sign of systematic velocity asymmetries across the jet at the 3$\sigma$ level, at any distance from the star. In the region close to the star ($\sim$ -0$\farcs$3 to -0$\farcs$6), we do see a consistent slope in radial velocity profile over the 3-5 spaxals across the jet.  Figure~\ref{fig3} shows that, at a distance of -0$\farcs$5 from the star, the peak signal-to-noise on the jet axis is 30. Interpolating this plot allows us to more precisely examine matching positions either side of the jet axis. Table \ref{tab2} give the results at -0$\farcs$5 from the source where the velocity is largest. We find differences in Doppler shift of 7$\pm$3 km~s$^{-1}$ (1$\sigma$) at a radius of 0$\farcs$07 (i.e. 10 AU) from the jet axis where the signal-to-noise drops to 19. Further from the axis, at 0$\farcs$1, the signal-to-noise drops even more (to 11) and the difference is 4$\pm$6 km~s$^{-1}$ (1$\sigma$). Hence, although it is possible that a signature of jet rotation in present in the innermost jet region, it is at most at the 2$\sigma$ level and thus not a conclusive detection. Unfortunately, the jet signal-to-noise for the anti-parallel slit was not high enough to verify this trend. The slope persists until -0$\farcs$6, just before the first jet knot, but not beyond. 

\begin{table}
\caption{\label{tab2}
Differences in Doppler Shift across the RY Tau jet at 0$\farcs$5 from star, based on interpolation of data in figure \ref{fig3}. Errors are at the 1$\sigma$ level.} 
\begin{center}
\begin{tabular}{@{}*{5}{l}}
\tableline
Radius from axis &SNR &$v_1$  &$v_2$ &$v_2$ - $v_1$\cr 
(arcsec) & &(km~s$^{-1}$)  &(km~s$^{-1}$) & (km~s$^{-1}$) \cr 
\tableline
0.07 &19 & -71$\pm$2  &-78$\pm$2  &-7$\pm$3 \cr
0.1 &11 & -72$\pm$4  &-77$\pm$4  &-5$\pm$6 \cr
\tableline
\end{tabular}
\end{center}
\end{table}

Jet radial velocities can be used to find the jet inclination angle, and hence the jet toroidal velocity. Radial velocities are in the range -60 to -110 km~s$^{-1}$. We note that the radial velocity is highest ($\sim$ -90 to -108 km~s$^{-1}$) in the region closest to the source ($\sim$ -0$\farcs$3), and at various points along the jet axis. Taking a typical value of -75 km~s$^{-1}$, and a proper motion of $\sim$ 165 km~s$^{-1}$ \citep{St-Onge2008}, we calculate an inclination angle with respect to the line of sight of 65.5$\degr$. Note that this rough estimate of the jet inclination from our observed radial velocities and the proper motions reported on large scales is compatible with the disk inclination on 100 AU scales of 70$\pm$5$\degr$ \citep{Isella2010}. We use this to find the jet toroidal velocity (section \ref{discussion}). 

\subsection{Disk Rotation} 

\begin{figure}
\begin{center}
\includegraphics[angle=-90, width=0.75\columnwidth]{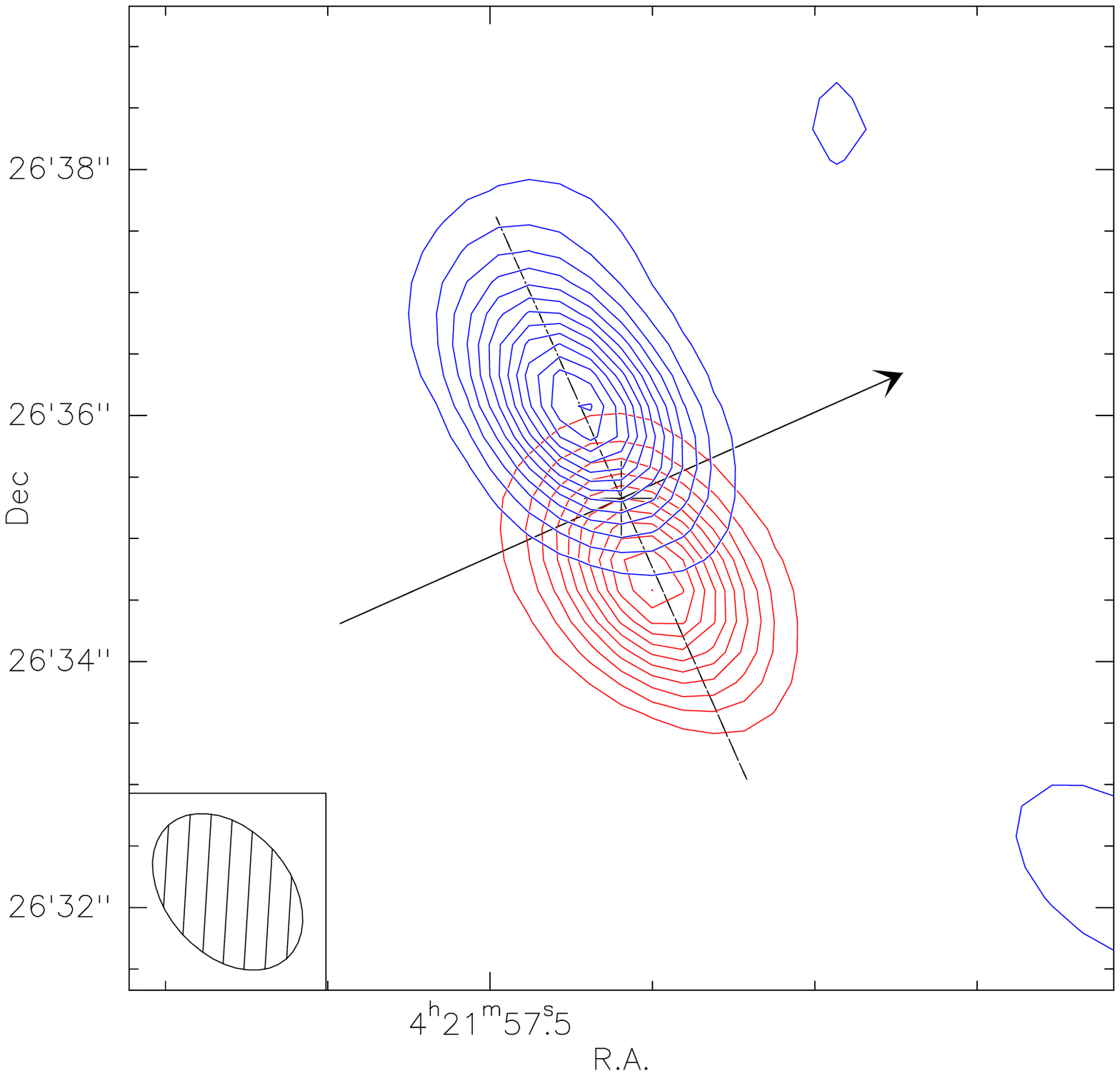} 
\caption{$^{12}$CO(2-1) channel maps from Plateau de Bure observations, illustrating the disk rotation sense. Blue and red contours show $^{12}$CO emission integrated from LSR velocities of -2.8 to -4.4 km~s$^{-1}$, and 9.5 to 11.1 km~s$^{-1}$, respectively.  Contours start at 3 $\sigma$. The beam size is shown as an insert. A clear velocity gradient is observed at PA=24$^{\circ}$, perpendicular to the black arrow which represents the approaching jet (PA= 294$\degr$). 
\label{fig5}
}
\end{center}
\end{figure}

\begin{figure}
\begin{center}
\includegraphics[ angle=-90, width=0.75\columnwidth]{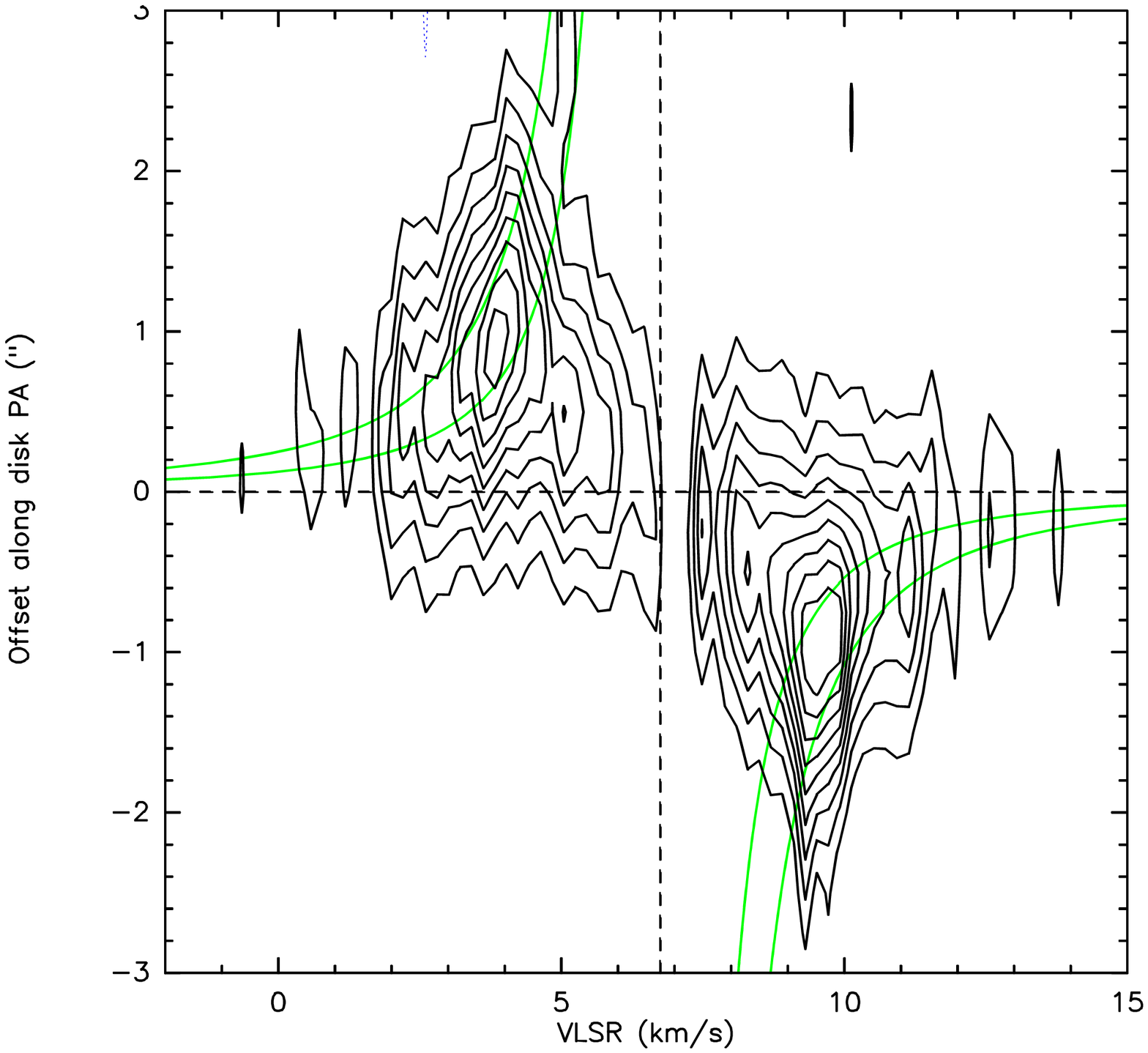} 
\caption{$^{12}$CO(2-1) position-velocity diagram along the disk PA, from Plateau de Bure observations. The two green curves show predictions for Keplerian rotation around a 1 and 2 M$_{\sun}$ star. The overlap of the contour peaks with the curves illustrates that the disk is in Keplerian otation. A disk inclination of 70$\degr$ was used, based on \citet{Isella2010}. The dashed line marks an estimate of the systemic velocity of 6.75 km~s$^{-1}$.
\label{fig6}
}
\end{center}
\end{figure}

Plateau de Bure observations of the RY Tau disk reveal detections of $^{12}$CO(2-1) emission from $v_{LSR}$ = 1.79 to 11.55 km~s$^{-1}$, i.e. covering a velocity range of 10.3 km~s$^{-1}$. Contour plots of the CO red and blue wing emissions, figure \ref{fig5}, reveal a clear velocity gradient along PA=24$\degr$, i.e. perpendicular to the jet axis and consistent with the disk PA derived by \citet{Isella2010}. The sense of the velocity gradient (blue-shifted emission to the north and red-shifted to the south) implies that the disk spin vector points along the axis of the approaching jet. A contour plot of the CO position-velocity diagram along PA=24$\degr$ shows that the  velocity field is consistent with Keplerian rotation around a 2 M$_{\sun}$ star, figure \ref{fig6}. The value of the systemic velocity suggested by the CO kinematics is estimated as $v_{sys}$ = 6.75 $\pm$ 0.2 km~s$^{-1}$. We use an inclination of the disk rotation axis to the line of sight of 70 $\pm$ 5 $\degr$ \citep{Isella2010}.  A more detailed study of the disk, via more complex modelling, is beyond the scope of this paper and is to be presented separately (Dougados et al. in preparation). 

\section{Discussion}\label{discussion} 

In this section, we examine the implications for jet launching, under a steady state assumption, of our derived upper limits on radial velocity differences across the jet. 

\subsection{Upper limits on jet toroidal velocity and angular momentum at 70 AU above the disk-plane}

While we do not detect any consistent rotation signature above the 3 $\sigma$ level in any of our transverse cuts of the jet, we can use our 3$\sigma$ upper limit on $(v_2-v_1)$ derived in section \ref{results} to determine upper limits on any possible toroidal velocity that may be present. 

In doing so, we note that an important consideration when attempting to constrain models for jet launching with signatures of jet toroidal velocities is that measured velocity differences across the jet always underestimate the true rotation profile, especially at small transverse distances from the jet axis. This is due to integration of the flow along the line of sight, combined with beam smearing \citep{Pesenti2004}. Correction for this depends on the spatial resolution of the observations, the distance from the jet axis where differences are measured, the jet velocity interval considered, and the jet ionisation profile, among other parameters. From the study of \citet{Pesenti2004}, we estimate that beam smearing effects can underestimate the toroidal velocity by up to a factor 2 at r=0$\farcs$07, and a factor 1.2 at r=0$\farcs$1. In order to minimise the beam smearing effects, we thus take the 3 $\sigma$ value at r=0$\farcs$1 to estimate the upper limit on $(v_2-v_1)$ to find an upper limit on the toroidal velocity. 

For a jet inclination angle of $i=$70$\degr$ to the line of sight, the 3 $\sigma$ value at r=0$\farcs$1 of 18 km~s$^{-1}$ translates to an upper limit on the toroidal velocity of $v_{\phi} = (v_2 - v_1)$ / (2$sin$($i$)) $<$ 10 km~s$^{-1}$ at a radius from the jet axis of $r$ = 14 AU and a distance above the disk-plane of $z$ = 70 AU. This implies a maximum value for specific angular momentum of $rv_{\phi}$=140 AU km~s$^{-1}$. 

\subsection{Implications for jet launching under a steady state, axisymmetric assumption} 
\label{implications}

In examining the implications of our toroidal velocity upper limit on jet launching models, we first consider that the steady state assumption is valid for the RY Tau jet at 70 AU from the source. Although shocks and wiggles are clearly present along the jet, they may only represent small perturbations in an underlying steady launching process. I.e. since the ambient medium has already been cleared by the jet in this evolved T Tauri system, we can be sure that the shocks seen in [\ion{Fe}{2}] are {\it internal} shocks where high-velocity jet material meets previously ejected lower velocity jet material. Hence, the observed radial velocities represent the {\it jet} radial velocity in the post-shock regions. Indeed, observed proper motions and analyses of line ratios in protostellar jets indicate shock speeds less than 25 \% of the flow velocity on scales of a few 100 AU  \citep{Lavalley-Fouquet2000}. Such shock amplitudes are consistent with the observed variation of radial velocities in the inner RY Tau jet, i.e. ($v_2 - v_1$) = 20 km~s$^{-1}$ compared to the mean of 85 km~s$^{-1}$. This also suggests that the observed flow velocities should be close to the mean ejection velocities from the disk. 

Toroidal velocity is a valuable parameter in differentiating between competing steady-state MHD launching mechanisms (e.g. \citealp{Bacciotti2002}; \citealp{Ferreira2006}; \citealp{Anderson2003}). Under the assumption of steady MHD ejection, a relation can be found between the specific angular momentum($r v_\phi$) and the poloidal velocity ($v_p$) for a given jet streamline. This relationship depends on one parameter: in the case of disk winds, the magnetic lever arm, $\lambda$, or launching radius, $r_0$, such that \citep{Ferreira2006, Anderson2003}: 
\begin{equation} 
\frac{rv_{\phi}v_p}{GM} = \lambda_{\phi}\sqrt{2\lambda_{\phi} - 3} 
\end{equation}
\begin{equation}
2rv_{\phi}\Omega_0 = v^2_p + 3\Omega^2_0r^2_0
\end{equation}
where $\Omega_0$ is the angular velocity at a radius in the disk-plane of $r_0$, and $\lambda_{\phi}$ is the effective lever arm at a given position from the driving source. Note that the value of $\lambda_{\phi}$ will remain lower than the true magnetic lever arm value, $\lambda$, until all angular momentum has been transferred to the matter. As noted in citealp{Ferreira2006}, this complete transfer may be reached only quite far from the star. For example, they show a solution with $\lambda$ = 13 which has $\lambda_{\phi}$ = 8 at as distance from the source of 50 AU. 
From these two equations, and assuming no pressure effects ($\beta$=0), we see that simultaneous knowledge of both $v_p$ and $v_{\phi}$ allows us to constrain both $r_0$ and $\lambda_{\phi}$. We stress that these derived relations are very general, and are valid for all types of super-Alfvenic, stationary, axisymmetric, self-collimated MHD jets, since they only express the conservation of total angular momentum and energy along a given streamlines. We refer to \citet{Ferreira2006} for details of how to derive these equations.  

\begin{figure}
\begin{center}
\includegraphics[ width=1\columnwidth]{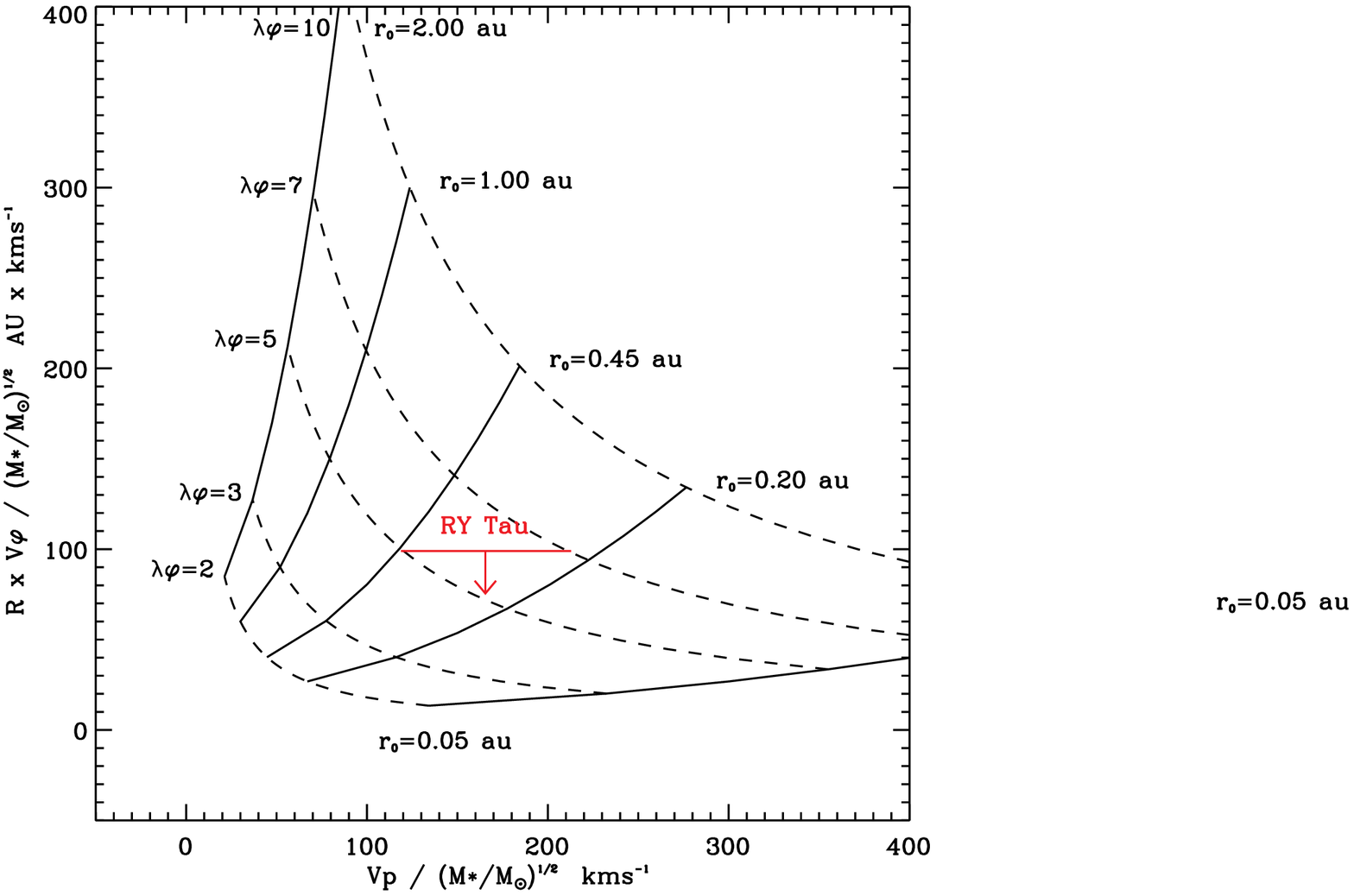} 
\caption{Comparison of launching mechanisms as a function of the relationship between jet poloidal and specific angular momentum (similar to \citet{Ferreira2006}). RY Tau is plotted according to our measurements. The RY Tau coordinates imply an upper limit on the jet launch radius in the disk of $r_0 <\le$ 0.45 AU, under the assumption that the jet is still in steady-state at 0$\farcs$5 (70 AU) from the star and a value of $\lambda_{\phi} \le$ 5. Recall that the range in $v_p$ comes from the range for the jet inclination angle of 70 $\pm$5$\degr$. 
\label{fig7}
}
\end{center}
\end{figure}

Figure~\ref{fig7} illustrates the parameter space of the $rv_\phi$ versus $v_p$ plane. Plot contours highlight values of $\lambda_{\phi}$ and $r_0$.  Only predictions for extended disk wind models are shown. The X-wind is a limiting case of these solutions with $r_0$=0.05 AU and $\lambda$=3 \citep{Agra-Amboage2009}. Stellar wind predictions fall below the lowest $r_0$ curve \citep{Ferreira2006}. 

A smaller toroidal velocity at a given radial distance from the jet axis implies smaller specific angular momentum, $rv_{\phi}$. Figure \ref{fig7} shows that for a given $v_p$, this then implies a smaller launching radius, $r_0$. Hence, we can use our derived upper limit on the specific angular momentum to derive an upper limit on the launching radius of the [\ion{Fe}{2}]-emitting jet from RY Tau. 

Using our velocity measurements, we plot RY Tau on the graph by \cite{Ferreira2006}. To do this, we calculate the magnitude of the jet poloidal velocity, (from the radial velocities of -71 to -78 km~s$^{-1}$ from table \ref{tab2} and jet inclination angle of $i$=70$\pm$5$\degr$) to be -168 to -301 km~s$^{-1}$, and take the stellar mass of $M_\star/M_\sun$ = 2. The resulting coordinate positioning indicates an upper limit on the launch radius in the disk, $r_0$$\le$ 0.45 AU, and a small magnetic lever arm parameter, $\lambda$=($r_A$/$r_0$)$^2$$\le$ 7. Note that our constraint on $r_0$ is only applicable for the part of the flow with radial velocities larger than the value at the radius from the jet axis of $r$=0$\farcs$1 and at 70 AU from the source. There are clearly data points with lower radial velocities in the outer borders of the jet, and this jet material could be originating from larger radii than $r_0 \le$0.45 AU. However, in our observations, these data are too faint to analyse. 

We briefly consider the implications of mass fluxes for the launching radius in the RY Tau system. In a steady MHD disk wind that carries away all of the accretion angular momentum flux from a Keplerian disk, there is a single inverse relation which has to hold between the magnetic lever arm parameter, $\lambda$, and the ratio of ejected to accreted mass-flux. We can then check that the derived $\lambda_{\phi}$ and $r_{0}$ parameters are compatible with the observed jet mass loss rate (\citealp{Ferreira2006}, equation 17): 
\begin{equation} 
\frac{\dot{M_{j}}}{\dot{M_{a}}} = \frac{1}{4(\lambda-1)} ln\frac{r_{out}}{r_{in}}   
\end{equation}
where $\dot{M_{j}}$ and $\dot{M_{a}}$ are the mass ejection and accretion rates respectively, and $r_{in}$ and $r_{out}$ are the inner and outer ejection radii on the disk plane. It is not a degeneracy of oversimplified self-similar models, but a basic requirement coming from energy and angular momentum conservation (see e.g. \citealp{Pudritz2007}). An estimate of the mass flux in the RY Tau blue-shifted atomic jet is between 0.16 and 2.6 $\times$ 10$^{-8}$ M$_{\odot}$ yr$^{-1}$ \citep{Agra-Amboage2011}, while accretion rate estimates range between 6.4-9 $\times$ 10$^{-8}$ M$_{\odot}$ yr$^{-1}$. This gives the ratio $\dot{M_{j}}/\dot{M_{acc}}$ = 0.02-0.4. If we take our constraint on $r_{out} = r_0 \le$ 0.45 AU, and assuming $r_{in}$ = 0.1 AU (i.e. close to co-rotation), then the (one-sided) mass flow ratio in RY Tau implies $\lambda \le$ 1.9-19.8. This is consistent with our derived constraint $\lambda$ $\le$ 7.  

\subsection{Implications for jet launching without the steady state, axisymmetric assumption}

We stress that our upper limits on $r_0$ and $\lambda_{\phi}$ are only valid if the jet is in steady state. Otherwise, the MHD invariants used to derive the formulae above would no longer be applicable. Indeed, the steady-state assumption may be uncertain, even at 0$\farcs$5 from the source, since our intensity map (fig \ref{fig2}) shows an emission knot as close as 0$\farcs$35 from the star which is associated with a velocity peak at -110 km~s$^{-1}$, indicating the occurrence of shocks very close to the source. Similarly, the detection of small jet wiggles beyond 0$\farcs$4 also makes the assumption of axisymmetry uncertain at this position. 

Recent numerical works have investigated the effect of shocks and jet propagation on rotation signatures in disk winds \citet{Fendt2011} show that magneto-hydrodynamic shocks occurring in an initially non-rotating jet with helical magnetic field configuration can transfer magnetic angular momentum (magnetic stress) into kinetic angular momentum (jet rotation). Conversely, \citet{Sauty2012} demonstrate that layers of a rotating disk wind may be temporarily counter-rotating due to initial variations in the flow velocity, without contradicting the MHD jet launching mechanism. Similarly, Staff et al. (2014, 2015) report 3D MHD simulations of disk wind propagation which show that kink instabilities can result in a signature of rotation in the outer jet layers in a direction opposite to that of the disk. If indeed time dependent effects are critical in structuring jets even as close to the star as investigated here (i.e. 70 AU), then our derived upper limit on angular momentum does not provide a reliable constraint on the launching radius and magnetic lever arm for RY Tau. Smaller spatial scales ($\sim$ 10 AU) should be investigated, where either the jet is steady and axisymmetric or where direct comparison with time-dependant 3D MHD numerical simulations including disk feedback becomes possible.

\section{Conclusions} 
\label{conclusions}

High resolution observations of jets close to their base are critical in constraining the jet ejection mechanism. Here, we present results for RY Tau, via an IFU dataset of high resolution both spatially and spectrally. We also present Plateau de Bure observations of the disk, which reveal a velocity field consistent with disk rotation, and we attempt to determine if we can observe a transfer of angular momentum from disk to jet. A radial velocity analysis of the jet revealed that there is no systematic difference across the jet at the 3 $\sigma$ level, i.e. no indication of jet rotation. Nevertheless, we can use the 3 $\sigma$ error bars to obtain a useful upper limit on the current jet toroidal velocity. For RY Tau, at 0$\farcs$5 (70 AU) from star and 0$\farcs$1 (14 AU) from axis, we find an upper limit of 10 km~s$^{-1}$ on the toroidal velocity. If we assume the jet is still in a steady-state at this distance from the star, the toroidal velocity would in turn provide an upper limit on the jet ejection radius in the disk plane of $r_0 \le$ 0.45 AU, and a magnetic lever arm parameter of $\le$ 5. (Recall that our constraint on $r_0$ does not hold for jet material of lower radial velocity in this region which may originate from a larger launching radius.) These results cannot differentiate between competing models but, strictly under the assumption of steady-state, the measurements can place important constraints on steady-state MHD models of jet launching. However, the steady-state assumption is challenged by our findings of multiple internal shocks starting as close as 0$\farcs$3 (50 AU) from the source, and a jet wiggle which develops further out. In this case, no firm constraint can be placed on the jet launch point since most of its angular momentum could be stored in magnetic form, rather than in rotation of matter. Although we do not reach the velocity accuracy anticipated, the variations within the IFU dataset make it clear that a three-dimensional datacube is critical to interpretations of velocity asymmetries across T Tauri jets. 

\section*{Acknowledgements}
Based on observations obtained at the Gemini Observatory, which is operated by the 
Association of Universities for Research in Astronomy, Inc., under a cooperative agreement 
with the NSF on behalf of the Gemini partnership: the National Science Foundation 
(United States), the National Research Council (Canada), CONICYT (Chile), the Australian 
Research Council (Australia), Minist\'{e}rio da Ci\^{e}ncia, Tecnologia e Inova\c{c}\~{a}o 
(Brazil) and Ministerio de Ciencia, Tecnolog\'{i}a e Innovaci\'{o}n Productiva (Argentina). 
Based on observations carried out with the IRAM Plateau de Bure Interferometer. IRAM is supported by INSU/CNRS (France), MPG (Germany) and IGN (Spain).
D.C. and C.D. wish to acknowledge the financial contributions of the School of Physics, University College Dublin, Ireland, and IPAG, Grenoble, France. 
We gratefully acknowledge the GEMINI Helpdesk team, in particular Emma Hogan and Marie Lemoine-Busserolle, for all support in ensuring the most accurate data calibration. We also thank Tracy Beck for her kind assistance in understanding the GEMINI pipeline. 

\bibliographystyle{apj}			

\end{document}